\documentclass[12pt]{article}

\usepackage[utf8]{inputenc}
\usepackage{geometry,setspace}
\usepackage{amsmath,amssymb,amsfonts,amsbsy,bm,bbm,latexsym,mathabx}
\usepackage{natbib,hyperref}
\usepackage{enumitem}
\usepackage{multirow,booktabs}
\usepackage{graphicx,epsfig,subcaption,xcolor}
\usepackage{algorithm,algpseudocode}
\usepackage{changes}

\newtheorem{theorem}{Theorem}
\newtheorem{lemma}{Lemma}
\newtheorem{remark}{Remark}
\newtheorem{corollary}{Corollary}

\def\D{\mathbf{D}}
\def\E{\mathbb{E}}
\def\G{\mathbf{G}}

\def\bbR{\mathbb{R}}

\def\V{\mathbb{V}}
\def\x{\mathbf{x}}
\def\X{\mathbf{X}}

\def\1{\mathbf{1}}

\def\bDelta{\boldsymbol{\Delta}}

\def\boxit#1{\vbox{\hrule\hbox{\vrule\kern6pt\vbox{\kern6pt#1\kern6pt}\kern6pt\vrule}\hrule}}

\title{DOD: Detection of outliers in high dimensional data with distance of distances}
\author{Seong-ho Lee$^{1}$, Yongho Jeon$^{2,3}$\thanks{corresponding author; email: \texttt{yhjeon@yonsei.ac.kr}}
\bigskip\\
$^1$Department of Statistics, University of Seoul, South Korea\\
$^2$Department of Statistics and Data Science, Yonsei University, South Korea\\
$^3$Department of Applied Statistics, Yonsei University, South Korea}
\date{}

\begin{document}
\onehalfspacing
\maketitle

\begin{abstract}
\noindent
Reliable outlier detection in high-dimensional data is crucial in modern science, yet it remains a challenging task. Traditional methods often break down in these settings due to their reliance on asymptotic behaviors with respect to sample size under fixed dimension. Furthermore, many modern alternatives introduce sophisticated statistical treatments and computational complexities. To overcome these issues, our approach leverages intuitive geometric properties of high-dimensional space, effectively turning the curse of dimensionality into an advantage. We propose two new outlyingness statistics based on observation's relational patterns with all other points, measured via pairwise distances or inner products. We establish a theoretical foundation for our statistics demonstrating that as the dimension grows, our statistics create a non-vanishing margin that asymptotically separates outliers from non-outliers. Based on this foundation, we develop practical outlier detection procedures, including a simple clustering-based algorithm and a distribution-free test using random rotations. Through simulation experiments and real data applications, we demonstrate that our proposed methods achieve a superior balance between detection power and false positive control, outperforming existing methods and establishing their practical utility in high-dimensional settings.
\end{abstract}



\clearpage
\doublespacing
\section{Introduction}\label{sec:intro}

The proliferation of high-dimensional data across scientific and industrial domains, from genomics and medical imaging to financial markets, has established outlier detection as a crucial analytical task. In these fields, outliers are often not merely noise to be discarded, but can represent the primary objects of interest, such as rare genetic variants associated with a disease, fraudulent financial transactions, or critical system failures \citep{chandola2009anomaly}. This task is particularly crucial in high-dimension, low-sample-size (HDLSS) settings, where the number of features $p$ vastly exceeds the number of observations $n$. In such scenarios, even a single outlier can cause serious distortions in statistical analysis, underscoring the need for robust and effective outlier detection methods.

Traditionally, a wide array of outlier detection methods were developed under the classical low-dimension, high-sample-size paradigm. These include methods based on distributional assumptions or approximations \citep{mcgill1978variations, ye2001anomaly}, density-based clustering \citep{ester1996density}, and nearest-neighbor heuristics \citep{breunig2000lof}. Many of these classical methods rely on metrics, such as the Mahalanobis distance, that summarize the data's multivariate distribution to detect observations that deviate from the norm. However, the performance of these metrics degrades severely in high dimensions due to the ``curse of dimensionality'' \citep{zimek2012survey}. The core of the problem lies in {their reliance on the large sample asymptotics under fixed dimension. For instance, for} the estimation of the covariance matrix, especially in HDLSS settings where $p \gg n$, the sample covariance matrix is singular and cannot be inverted, or its estimate is subject to high variability. This statistical and numerical instability causes the collapse of classical metrics, rendering many well-established outlier detection methods ineffective or entirely inapplicable for high-dimensional data.

In response to these challenges, a new generation of methods designed specifically for high-dimensional data has emerged. These methods include approaches based on measuring local density variation \citep{papadimitriou2003loci}, using angles instead of distances \citep{kriegel2008angle}, and developing robust versions of classical methods. For instance, \cite{filzmoser2008outlier} proposed a reweighting method using principal components, while \cite{ro2015outlier} improved the Mahalanobis distance by using the minimum covariance determinant. More recently, \cite{chung2021subspace} introduced a metric based on a distance to hyperplane, and devised a two-stage procedure that conducts a hypothesis test for each outlier candidate. A key advantage of these methods is that they can operate without relying on{the large sample asymptotics}. However, this advantage comes with its own cost; many of these modern alternative methods require sophisticated statistical treatments thus computationally complex, along with careful tuning parameter settings.

To overcome these issues, we propose a new outlier detection method that is computationally simple while theoretically grounded. Our approach draws inspiration from the concept of distance vector clustering introduced by \cite{terada2013clustering}, which was shown to be an efficient alternative to methods based on the maximal data piling direction \citep{ahn2010maximal, ahn2012clustering}, as it discriminates groups based on metrics that are simple yet effectively reflect both mean and variance differences in different groups. Instead of relying on complex statistical treatments, we innovate this simple concept to devise an outlyingness statistic for individual observations by leveraging intuitive geometric properties of high-dimensional space. The core insight is that outliers exhibit a relational pattern with respect to all other data points that is fundamentally different from that of non-outliers (or inliers). We capture this characteristic by adopting the concepts of \textit{Distance of Distances} (DOD) and \textit{Distance of inner products in a Gram matrix} (DOG). These concepts lead to two new outlyingness measure statistics which quantify how much an observation's entire profile of pairwise relationships deviates from the typical profile of non-outlying points.

The contributions of this paper are three-fold. First, we establish a theoretical foundation for our statistics, demonstrating that as the dimension $p$ grows, the statistics create a non-vanishing asymptotic margin between outliers and non-outliers. Second, based on this theoretical foundation, we develop a set of practical outlier detection procedures, including a simple clustering-based algorithm, and data-driven non-parametric tests based on random rotations \citep{blaser2016random}, a technique effectively used in \cite{chung2021subspace}. Third, we demonstrate through simulation experiments and applications to two real datasets, a microarray gene expression dataset and a human face image dataset, that our methods achieve a superior balance between high detection power and stringent false positive control compared to existing methods.

The rest of the paper is organized as follows. Section \ref{sec:statistic} introduces the proposed statistics and their theoretical properties. Section \ref{sec:test} details the outlier detection procedures. Sections \ref{sec:sim} and \ref{sec:real} present the numerical results from simulation experiments and real data applications, respectively. Finally, Section \ref{sec:con} concludes the paper.


\section{Proposed statistics and theoretical properties}\label{sec:statistic}
\subsection{Proposed statistics}

\begin{figure}
    \centering
    \caption{Illustration of the proposed outlier detection statistic. Panel (\ref{sub@fig:data}) shows a 2D projection of a simulated dataset containing two outliers (7 and 17). Panel (\ref{sub@fig:bDelta}) shows the heatmap of $\bDelta_{\rm D}$. Panel (\ref{sub@fig:t}) shows the barplot of $t_i^{\rm (D)}$, which is markedly larger for the outliers.}
    \label{fig:illustration}

    \begin{subfigure}{0.275\textwidth}
        \centering
        \includegraphics[width=\textwidth]{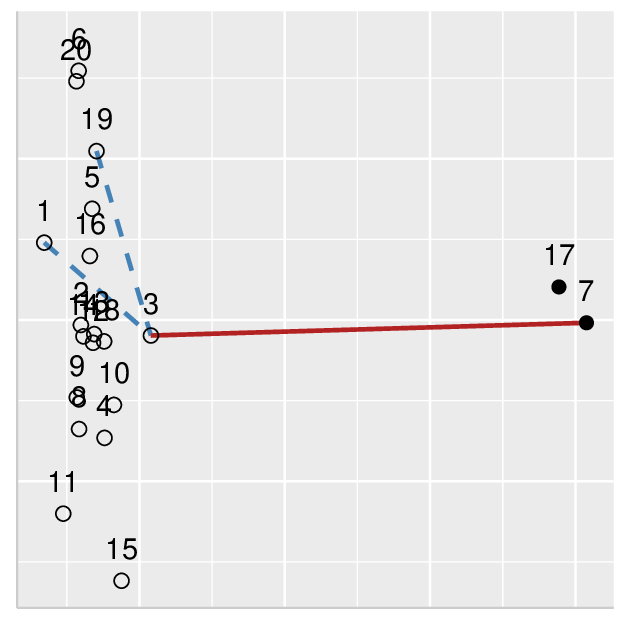}
        \caption{Data scatterplot}
        \label{fig:data}
    \end{subfigure}
    \hfill
    \begin{subfigure}{0.275\textwidth}
        \centering
        \includegraphics[width=\textwidth]{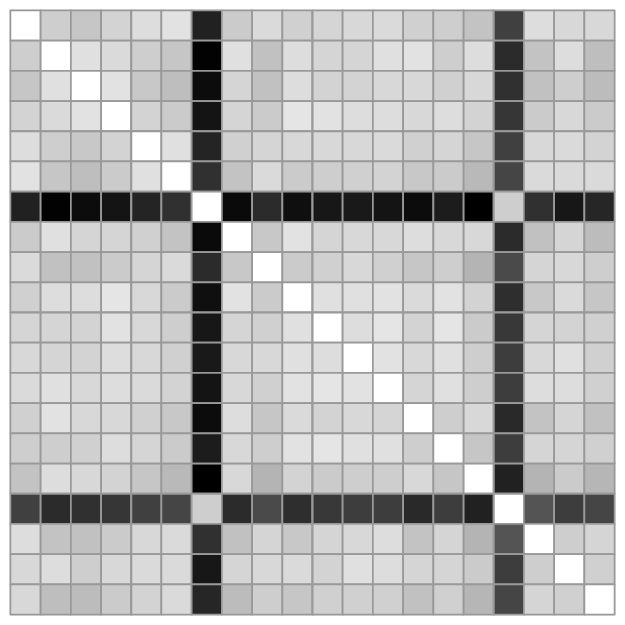}
        \caption{Heatmap of $\bDelta_{\rm D}$}
        \label{fig:bDelta}
    \end{subfigure}
    \hfill
    \begin{subfigure}{0.35\textwidth}
        \centering
        \includegraphics[trim=0pt 30pt 0pt 0pt, width=\textwidth]{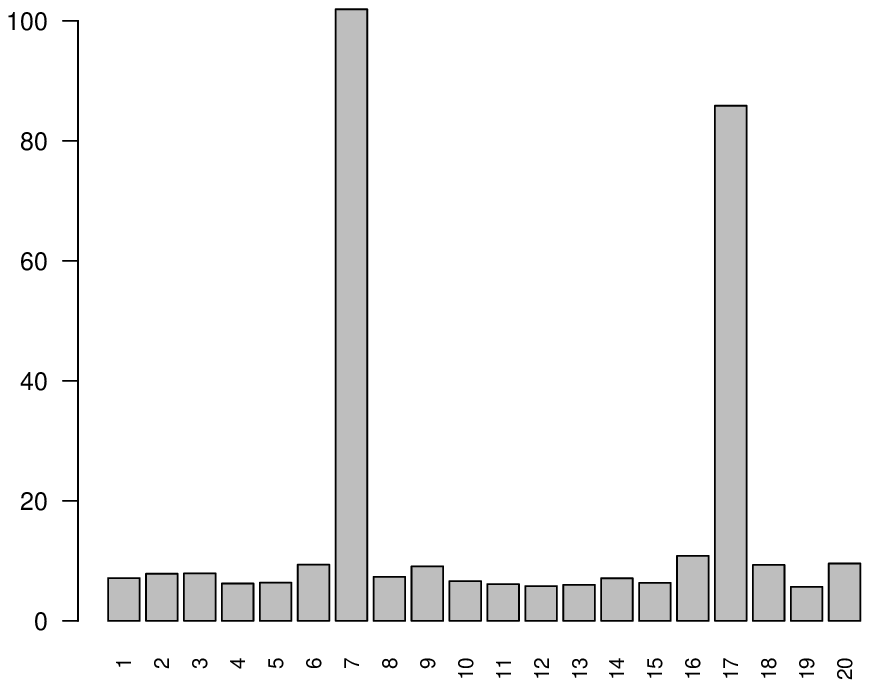}
        \caption{Barplot of $t_i^{\rm (D)}$}
        \label{fig:t}
    \end{subfigure}

\end{figure}

A core motivation for our proposed statistics stems from the observation of how outliers manifest in pairwise relationship matrices. Figure~\ref{fig:illustration} illustrates this phenomenon based on a simulated dataset with dimension $p=1000$ and sample size $n=20$, containing two designated outliers (the 7th and 17th observations). In the 2D projection shown in Figure~\ref{fig:illustration}(\ref{sub@fig:data}), a key visual takeaway is that the length of the solid line, representing the dissimilarity between an inlier and an outlier, is substantially greater than that of the dashed line, representing the dissimilarity between two inliers.

By aggregating this relational information for all pairs, we construct an $n\times n$ matrix $\bDelta_{\rm D}$ visualized as a heatmap in Figure~\ref{fig:illustration}(\ref{sub@fig:bDelta}), which will be elaborated in a sequel. Large dissimilarities between observations appear as dark entries in the heatmap. These dark entries form distinct, high-magnitude columns that correspond to the outliers. This clear pattern demonstrates that an outlier's relational profile is profoundly different from that of a non-outlier, which leads to a critical insight: the column-wise median of this matrix can serve as a robust baseline for the typical relational pattern, and consequently, a substantial deviation from this baseline can serve as a strong indicator of outlyingness.

This phenomenon directly underpins the design of our proposed statistics $t_i^{\rm (D)}$ and $t_i^{\rm (G)}$. We now elaborate our first proposed statistic $t_i^{\rm (D)}$. Let $\X\in\bbR^{n\times p}$ be a centered data matrix with its $i$th row denoted as $\x_i\in\bbR^{p}$. We start by computing the distance matrix $\D\in\bbR^{n\times n}$, where $[\D]_{i,j}=d(\x_i,\x_j)$. For ease of presentation, we use the Euclidean distance for $d(\cdot,\cdot)$. Based on this matrix, we construct the matrix of distances-of-distances, $\bDelta_{\rm D}\in\bbR^{n\times n}$ with elements $[\bDelta_{\rm D}]_{i,j}=\delta_{ij}^{\rm (D)}=\sqrt{\sum_{k \neq i,j}([\D]_{i,k}-[\D]_{j,k})^2}$. $\delta_{ij}^{\rm (D)}$ measures the dissimilarity between the distance patterns of $\x_i$ and $\x_j$ relative to all other observations. Following this construction, we define our first statistic $t_i^{\rm (D)}$ as the Euclidean distance between the $i$th row of $\bDelta_{\rm D}$ and the column-wise median vector of $\bDelta_{\rm D}$. The column-wise median vector serves as a robust representation of the typical pattern of non-outlying points. Formally,
\[
t_i^{\rm (D)} = \sqrt{\sum_{j=1}^{n}\left\{\delta_{ij}^{\rm (D)}-\widetilde{\delta}_{.j}^{\rm (D)}\right\}^2},
\]
where $\widetilde{\delta}_{.j}^{\rm (D)}=\text{median}\{\delta_{ij}^{\rm (D)}:i=1,\dots,n\}$.

Similarly, we propose a second statistic $t_i^{\rm (G)}$ based on inner products, which captures different aspects of dissimilarity from $t_i^{\rm (D)}$. We first compute the inner product matrix $\G\in\bbR^{n\times n}$, with elements $[\G]_{i,j}=\langle\x_i,\x_j\rangle$. We focus on the dot product. Analogous to $\bDelta_{\rm D}$, we construct the matrix of distances-of-inner-products $\bDelta_{\rm G}\in\bbR^{n\times n}$ with elements $[\bDelta_{\rm G}]_{i,j}=\delta_{ij}^{\rm (G)}=\sqrt{\sum_{k \neq i,j}([\G]_{i,k}-[\G]_{j,k})^2}$. This quantity $\delta_{ij}^{\rm (G)}$ measures the dissimilarity in inner product patterns between $\x_i$ and $\x_j$. Our second statistic $t_i^{\rm (G)}$ is defined as the Euclidean distance between the $i$th row of $\bDelta_{\rm G}$ and its column-wise median vector:
\[
t_i^{\rm (G)} = \sqrt{\sum_{j=1}^{n}\left\{\delta_{ij}^{\rm (G)}-\widetilde{\delta}_{.j}^{\rm (G)}\right\}^2},
\]
where $\widetilde{\delta}_{.j}^{\rm (G)}=\text{median}\{\delta_{ij}^{\rm (G)}:i=1,\dots,n\}$.

To illustrate the consequence of this design, Figure~\ref{fig:illustration}(\ref{sub@fig:t}) displays a barplot of the proposed statistic $t_i^{\rm (D)}$ computed under the same simulated dataset for Figures~\ref{fig:illustration}(\ref{sub@fig:data}) and \ref{fig:illustration}(\ref{sub@fig:bDelta}). Consistent with the patterns of the heatmap in Figure~\ref{fig:illustration}(\ref{sub@fig:bDelta}), the statistics corresponding to the two outliers---the 7th and 17th observations--- are markedly larger than those for the remaining non-outlying points. This stark separation of magnitudes provides an empirical demonstration of the utility of our proposed statistics in outlier detection. The theoretical justification for this phenomenon, which guarantees a clear margin between the outlier and non-outlier statistics in high-dimensional settings, is established in Theorems~\ref{th:t} and~\ref{th:margin} in the following section.


\subsection{Theoretical properties}

To rigorously validate the effectiveness of our proposed statistics $t_i^{\rm (D)}$ and $t_i^{\rm (G)}$ under high-dimensional settings, we now establish the theoretical properties of the proposed statistics under a set of assumptions. Let $\mathbf{x}^{\rm (I)}=[X_1^{\rm (I)},\dots,X_p^{\rm (I)}]^\top\in\bbR^p$ be the random vector representing non-outliers and $\mathbf{x}^{\rm (O)}=[X_1^{\rm (O)},\dots,X_p^{\rm (O)}]^\top\in\bbR^p$ be the random vector representing outliers. Following the framework of \cite{hall2005geometric}, a common approach in high-dimensional asymptotic studies, we assume the following conditions.

\begin{enumerate}[label=(H\arabic*)]
    \item\label{con:4th}
    The fourth moments of the entries of the sample vectors are uniformly bounded.
    \item\label{con:mean}
    $\lim_{p\to\infty}\frac{1}{p}\sum_{k=1}^{p} \E\{X_k^{\rm (I)}\}^2 = \mu_I^2$ and $\lim_{p\to\infty}\frac{1}{p}\sum_{k=1}^{p} \E\{X_k^{\rm (O)}\}^2 = \mu_O^2$.
    \item\label{con:var}
    $\lim_{p\to\infty}\frac{1}{p}\sum_{k=1}^{p} \V\{X_k^{\rm (I)}\} = \sigma_I^2$ and $\lim_{p\to\infty}\frac{1}{p}\sum_{k=1}^{p} \V\{X_k^{\rm (O)}\} = \sigma_O^2$.
    \item\label{con:meandiff}
    $\lim_{p\to\infty}\frac{1}{p}\sum_{k=1}^{p} \left[\E\{X_k^{\rm (I)}\}-\E\{X_k^{\rm (O)}\}\right]^2 = \delta^2$.
    \item\label{con:rhomixing}
    For all random vectors, there exists a permutation of entries such that
    the sequence of the variables are $\rho$-mixing for functions that are dominated by quadratics.
\end{enumerate}

These conditions provide a foundation to analyze the limiting behavior of the proposed statistics. Specifically, Conditions \ref{con:mean} and \ref{con:var} ensure that the per-feature mean squared expectation and variance of both non-outliers and outliers converge to fixed values as the dimension $p$ grows. This allows for a stable characterization of each group. Condition \ref{con:meandiff} is also crucial in that it formalizes the separation between the non-outlier and outlier clusters, ensuring that the squared mean difference between the two groups does not vanish in the high-dimensional limit.

To further provide a theoretical support for the proposed statistics, we state the asymptotic behavior of the constituent quantities of $\bDelta_{\rm D}$ and $\bDelta_{\rm G}$ as $p \to \infty$. The following lemma, a corrected and restated version of the result from \cite{terada2013clustering} under the assumptions from \cite{hall2005geometric}, provides the asymptotic limits for the pairwise differences in distances and inner products.

\begin{lemma}\label{lem:terada}
Under Conditions \ref{con:4th}--\ref{con:rhomixing}, we have the following results as $p\to\infty$:
\begin{enumerate}[label=(\roman*)]
    \item If either $\mathbf{x}_i, \mathbf{x}_j \sim \mathbf{x}^{\rm (I)}$ or $\mathbf{x}_i, \mathbf{x}_j \sim \mathbf{x}^{\rm (O)}$,
    \begin{align*}
        \frac{1}{\sqrt{p}} ([\D]_{i,k}-[\D]_{j,k}) &\xrightarrow{p} 0,\\
        \frac{1}{p} ([\G]_{i,k}-[\G]_{j,k}) &\xrightarrow{p} 0.
    \end{align*}

    \item If $\mathbf{x}_i \sim \mathbf{x}^{\rm (I)}$ and $\mathbf{x}_j \sim \mathbf{x}^{\rm (O)}$,
    \begin{align*}
        \frac{1}{\sqrt{p}} ([\D]_{i,k}-[\D]_{j,k}) &\xrightarrow{p}
        \begin{cases}
            \sqrt{2}\sigma_\text{I}-\sqrt{\sigma_\text{I}^2+\sigma_\text{O}^2+\delta^2} := \alpha_{\rm D} & \text{if } \mathbf{x}_k \sim \mathbf{x}^{\rm (I)}, \\
            \sqrt{\sigma_\text{I}^2+\sigma_\text{O}^2+\delta^2}-\sqrt{2}\sigma_\text{O} := \beta_{\rm D} & \text{if } \mathbf{x}_k \sim \mathbf{x}^{\rm (O)}.
        \end{cases} \\
        \frac{1}{p} ([\G]_{i,k}-[\G]_{j,k}) &\xrightarrow{p}
        \begin{cases}
            \frac{\mu_\text{I}^2-\mu_\text{O}^2+\delta^2}{2} := \alpha_{\rm G} & \text{if } \mathbf{x}_k \sim \mathbf{x}^{\rm (I)}, \\
            \frac{\mu_\text{I}^2-\mu_\text{O}^2-\delta^2}{2} := \beta_{\rm G} & \text{if } \mathbf{x}_k \sim \mathbf{x}^{\rm (O)}.
        \end{cases}
    \end{align*}

    \item We have $\alpha_{\rm D}=\beta_{\rm D}=0$ if and only if $\sigma_\text{I}^2=\sigma_\text{O}^2$ and $\delta^2=0$. Also, $\alpha_{\rm G}=\beta_{\rm G}=0$ if and only if $\mu_\text{I}^2=\mu_\text{O}^2$ and $\delta^2=0$.
\end{enumerate}
\end{lemma}

Lemma \ref{lem:terada} suggests that these pairwise differences converge to distinct, non-zero values depending on whether the observations involved are non-outliers or outliers, which is a critical property for our statistics to effectively differentiate the two groups. Specifically, Lemma \ref{lem:terada}(iii) underscores the distinct characteristics of the two statistics; the distance-based measure is primarily sensitive to the discrepancy in population variances $\sigma_\text{I}^2$ and $\sigma_\text{O}^2$, while the inner-product-based measure captures the difference in squared norms $\mu_\text{I}^2$ and $\mu_\text{O}^2$.

Based on the asymptotic behaviors established in Lemma \ref{lem:terada}, we now present the main theoretical result concerning our proposed statistics. The following theorem formally demonstrates that our statistics can effectively distinguish between non-outliers and outliers in the high-dimensional setting. Its proof is provided in the supplementary material.

\begin{theorem}\label{th:t}
Under Conditions \ref{con:4th}--\ref{con:rhomixing}, we have the following results as $p \to \infty$:
\begin{enumerate}[label=(\roman*)]
    \item For a non-outlier $\x_i \sim \x^{\rm (I)}$, the scaled statistics converge to zero in probability:
    \begin{align*}
        \frac{1}{\sqrt{p}} \, t_i^{\rm (D)} \xrightarrow{p} 0 \quad \text{and} \quad \frac{1}{p} \, t_i^{\rm (G)} \xrightarrow{p} 0.
    \end{align*}
    \item For an outlier $\x_i \sim \x^{\rm (O)}$, the scaled statistics converge to constants in probability:
    \begin{align*}
        \frac{1}{\sqrt{pn}} \, t_i^{\rm (D)} &\xrightarrow{p} \sqrt{(n-n_{\rm out}-1)\alpha_{\rm D}^2 + (n_{\rm out}-1)\beta_{\rm D}^2} &:= \gamma_{\rm D}, \\
        \frac{1}{p\sqrt{n}} \, t_i^{\rm (G)} &\xrightarrow{p} \sqrt{(n-n_{\rm out}-1)\alpha_{\rm G}^2 + (n_{\rm out}-1)\beta_{\rm G}^2} &:= \gamma_{\rm G}.
    \end{align*}
\end{enumerate}
\end{theorem}

\begin{remark}[Individual Distinction]
Theorem \ref{th:t} demonstrates a distinction in the asymptotic behavior of the proposed statistics for non-outliers versus outliers. For any non-outlying observation, the scaled statistic is asymptotically negligible, as its value vanishes toward zero in the high-dimensional limit. In contrast, the scaled statistic for an outlier captures a signal of its anomalous nature, converging to a positive constant. This divergent limiting behavior provides a theoretical support for their distinction by adopting our proposed statistics.
\end{remark}

\begin{remark}[Enhanced Detection from Sample Size $n$ and Dimensionality $p$]
The performance of our proposed statistics is enhanced by both sample size $n$ and dimensionality $p$. Firstly, a larger sample size $n$ directly magnifies the statistic $t_i^{\rm (D)}$ for outliers, while leaving it unchanged for non-outliers. Specifically, for an outlier, the magnitude of $t_i^{\rm (D)}$ grows linearly with $n$ since $t_i^{\rm (D)} \approx\sqrt{pn}~\gamma_{\rm D}$ and $\gamma_{\rm D}~\propto~\sqrt{n}$. In contrast, for a non-outlier, its magnitude $o_p(\sqrt{p})$ is independent of $n$. This creates a widening gap between the outlier and non-outlier statistics as the sample size increases, thereby strengthening detection power.

Secondly, our method leverages high dimensionality. It relies on the convergent behavior of pairwise distances and inner products in high-dimensional spaces, where large $p$ ensures the asymptotic stabilization described in the conditions and theorems. This dimension leveraging effectively turns the classic curse of dimensionality into an advantage for outlier detection, making our proposed statistics useful for high-dimensional data.
\end{remark}

Building upon the individual convergence properties shown in Theorem \ref{th:t}, we now advance to a stronger, collective statement. While the previous theorem guarantees that the individual statistic for any non-outlier vanishes while that for an outlier remains large, it yet does not preclude the possibility of overlap between the two populations. The next theorem resolves this issue by proving that a non-vanishing margin indeed exists, separating the entire set of outliers from the set of non-outliers. Its proof is provided in the supplementary material.

\begin{theorem}\label{th:margin}
Let $\mathcal{I}$ and $\mathcal{O}$ be the index sets for non-outliers and outliers, respectively. Under Conditions \ref{con:4th}--\ref{con:rhomixing}, the gap between the scaled outlier and non-outlier statistics converges to constants in probability as $p \to \infty$:
\begin{align*}
    \min_{i \in \mathcal{O}} \frac{t_i^{\rm (D)}}{\sqrt{pn}} - \max_{i \in \mathcal{I}} \frac{t_i^{\rm (D)}}{\sqrt{pn}} &\xrightarrow{p} \gamma_{\rm D}, \\
    \min_{i \in \mathcal{O}} \frac{t_i^{\rm (G)}}{p\sqrt{n}} - \max_{i \in \mathcal{I}} \frac{t_i^{\rm (G)}}{p\sqrt{n}} &\xrightarrow{p} \gamma_{\rm G}.
\end{align*}
\end{theorem}

\begin{corollary}\label{cor:margin}
Under Conditions \ref{con:4th}--\ref{con:rhomixing}, the gap between the scaled outlier and non-outlier statistics is bounded strictly above zero in probability as $p\to\infty$:
\begin{align*}
    \lim_{p\to\infty}\Pr\left\{\min_{i \in \mathcal{O}} \frac{t_i^{\rm (D)}}{\sqrt{pn}} - \max_{i \in \mathcal{I}} \frac{t_i^{\rm (D)}}{\sqrt{pn}} >0\right\}&=1 \quad \text{if } \sigma_{\rm I}^2\neq\sigma_{\rm O}^2 \text{ or }\delta^2\neq0, \\
    \lim_{p\to\infty}\Pr\left\{\min_{i \in \mathcal{O}} \frac{t_i^{\rm (G)}}{p\sqrt{n}} - \max_{i \in \mathcal{I}} \frac{t_i^{\rm (G)}}{p\sqrt{n}} >0\right\}&=1 \quad \text{if } \mu_{\rm I}^2\neq\mu_{\rm O}^2 \text{ or }\delta^2\neq0.
\end{align*}
\end{corollary}

\begin{remark}[Existence of a Separation Margin]
Theorem \ref{th:margin} and Corollary \ref{cor:margin} provide a stronger theoretical guarantee for our statistics' outlier detection performance. It demonstrates that as the dimension grows, the two groups become perfectly separated; the smallest scaled statistic from the outlier group becomes strictly greater than the largest scaled statistic from the non-outlier group. This result offers a justification for distinguishing outliers, as a clear margin emerges between the two populations. The existence of this non-vanishing separation margin $\gamma_{\rm D}$ (or $\gamma_{\rm G}$) ensures that the detection capability of the proposed statistics is reliable in high-dimensional settings.
\end{remark}


\section{Proposed outlier detection procedure}\label{sec:test}

\subsection{Detection via clustering}

Our theoretical results provide a foundation for a practical detection procedure. The key insight stems from Theorem \ref{th:margin}, which guarantees that as the dimension $p$ grows, a non-vanishing margin emerges between the scaled statistics of non-outliers and outliers. This asymptotic separability is the cornerstone of our proposed procedure, as it effectively transforms the complex, high-dimensional outlier detection problem into a much simpler, one-dimensional clustering task performed on the set of statistics $\{t_i\}_{i=1}^n$.

Leveraging this theoretical guarantee, we propose a straightforward outlier detection procedure via clustering. The procedure begins by computing outlyingness statistics, either $t_i^{\rm (D)}$ or $t_i^{\rm (G)}$, for each observation $\mathbf{x}_i$ in the dataset. Subsequently, a standard clustering algorithm is applied to partition these $n$ statistics into two distinct groups. Given the clear separation shown by Theorem \ref{th:margin}, a simple algorithm such as k-means is sufficient to effectively distinguish the two populations.

The final step is to label the two clusters and validate their separation. The cluster with the larger mean statistic is designated as the potential outlier group, $C_{\text{out}}$. To avoid the pitfall of wrongly declaring this cluster as outliers in outlier-free scenarios, we validate the separation between the groups. We compute the gap defined as $g = \min_{i \in C_{\text{out}}} t_i - \max_{j \in C_{\text{in}}} t_j$, where $C_{\text{in}}$ is the non-outlier cluster. The members of $C_{\text{out}}$ are then declared as outliers only if this gap exceeds a predefined gap threshold $c>0$ and if the cluster's size $|C_{\text{out}}|$ is less than a specified proportion of the total sample size $n\alpha$. The parameter $\alpha \in (0, 0.5)$ represents the maximum proportion of outliers, thus serves as a tuning parameter {controling the maximum false positive rate (FPR)}. Otherwise, we conclude that no distinct group of outliers exists and return an empty set. This procedure is summarized in Algorithm \ref{alg:clustering}.

\begin{algorithm}[t]
\caption{Outlier Detection via Clustering}\label{alg:clustering}
\begin{algorithmic}[1]
    \Statex \textbf{Input}: Centered data matrix $\mathbf{X} \in \mathbb{R}^{n \times p}$, maximum {FPR} $\alpha$, gap threshold $c$.
    \Statex \textbf{Output}: Index set of declared outliers $\widehat{\mathcal{O}}$.
    \vspace{0.5em}
    
    \State Compute the statistics $\{t_1, \dots, t_n\}$ from $\mathbf{X}$.
    
    \State Partition $\{t_1, \dots, t_n\}$ into two clusters $C_1$ and $C_2$ using a clustering algorithm.
    
    \State Identify the potential outlier cluster $C_{\text{out}}$ as the cluster with the larger mean of the statistic, and $C_{\text{in}}$ as the other.
    
    \State Compute the gap: $g = \min_{i \in C_{\text{out}}} t_i - \max_{j \in C_{\text{in}}} t_j$.
    
    \If{$|C_{\text{out}}| \leq n\alpha$ and $g > c$}
        \State $\widehat{\mathcal{O}} \leftarrow \{i \mid i \in C_{\text{out}}\}$.
    \Else
        \State $\widehat{\mathcal{O}} \leftarrow \emptyset$.
    \EndIf
    \State \textbf{Return} $\widehat{\mathcal{O}}$.
\end{algorithmic}
\end{algorithm}


\subsection{Detection via random rotation}

As an alternative to clustering for outlier detection, we propose a non-parametric testing procedure based on random rotation \citep{blaser2016random}. Random rotation is a data perturbation technique where, for a data $\mathbf{X}$, a rotated version $\mathbf{X}^* = \mathbf{H}\mathbf{X}$ is generated by pre-multiplying a randomly sampled rotation matrix $\mathbf{H}$. This allows us to generate a reference or ``null" distribution for a test statistic directly from the observed data, providing a distribution-free, data-driven decision boundary for hypothesis testing. Thus, instead of relying on a predefined gap threshold $c$ as Algorithm \ref{alg:clustering}, a new proposed procedure will provide a data-driven threshold.

The theoretical justification for our proposal is grounded in the properties of the left-spherical distribution family \citep{chung2021subspace}. Let us establish a null hypothesis $H_0$ that the non-outlier data follows a left-spherical distribution. Under this hypothesis, the non-outlier data distribution is invariant to pre-multiplication by any orthogonal matrix $\mathbf{H}$. Rotating the entire dataset by pre-multiplying $\mathbf{H}$ to $\mathbf{X}$, we can simulate the distribution of test statistics under the null hypothesis, as the rotation randomizes observation-specific quantities while preserving the distribution of the entire dataset. It is worthwhile to note that our proposed statistics $t_i^{\rm (D)}$ and $t_i^{\rm (G)}$ are dependent on the relative arrangement of the observations therefore rotation-variant, making them well-suited for this procedure.

Specifically, the random rotation test for outlier detection can be implemented in two different ways, with the second offering superior statistical properties. A first, straightforward implementation of the random rotation test involves creating a null distribution by pooling all statistics from the rotated data. The procedure begins by computing the statistics $\{t_1, \dots, t_n\}$ for the original data. Subsequently, a number of rotated datasets $\mathbf{X}_b~(b=1,\dots,B)$ are generated, and the statistics $\{t_{1,b}, \dots, t_{n,b}\}$ are computed for each. All $n \times B$ of these statistics are then aggregated into a single empirical null distribution $\mathcal{T}_{\text{null}}$. The critical value $c_\alpha$ is then determined by the $(1-\alpha)$th quantile of this distribution. Finally, we declare the $i$th observation as an outlier if $t_i>c_\alpha$. This procedure is detailed in Algorithm \ref{alg:random_rotation_pooled}.

\begin{algorithm}[t]
\caption{Outlier Detection via Random Rotation}
\label{alg:random_rotation_pooled}
\begin{algorithmic}[1]
    \Statex \textbf{Input}: Centered data matrix $\mathbf{X} \in \mathbb{R}^{n \times p}$, {maximum FPR} $\alpha$, number of rotations $B$.
    \Statex \textbf{Output}: Index set of declared outliers $\widehat{\mathcal{O}}$.
    \vspace{0.5em}
    
    \State Compute the statistics $\{t_1, \dots, t_n\}$ from $\mathbf{X}$.
    \State Initialize an empty set for the null distribution: $\mathcal{T}_{\text{null}} \leftarrow \emptyset$.
    
    \For{$b = 1$ to $B$}
        \State Generate a random orthogonal matrix $\mathbf{H}_b$.
        \State Compute the rotated data matrix $\mathbf{X}_b = \mathbf{H}_b\mathbf{X}$.
        \State Compute the statistics $\{t_{1,b}, \dots, t_{n,b}\}$ from $\mathbf{X}_b$.
        \State Update the null distribution: $\mathcal{T}_{\text{null}} \leftarrow \mathcal{T}_{\text{null}} \cup \{t_{1,b}, \dots, t_{n,b}\}$.
    \EndFor
    
    \State Determine the critical value $c_\alpha$ as the $(1-\alpha)$th quantile of $\mathcal{T}_{\text{null}}$.
    \State Identify the outlier index set $\widehat{\mathcal{O}} = \{i \mid t_i > c_\alpha\}$.
    \State \textbf{Return} $\widehat{\mathcal{O}}$.
\end{algorithmic}
\end{algorithm}

While intuitive, this method fails to account for the multiple comparisons problem inherent in testing $n$ hypotheses simultaneously. Consequently, the probability of making at least one false discovery is not controlled at the nominal level $\alpha$, potentially leading to an inflated number of false positives. To address this shortcoming, we further propose a procedure that controls the Family-Wise Error Rate (FWER). The FWER is the probability of making one or more false discoveries, thus controlling it provides a much stronger guarantee of statistical validity.

This is achieved by constructing the null distribution of the maximum statistic from each rotated data $t_{\max, b} = \max_i \{t_{i,b}\}$. The collection of these maximums forms an empirical null distribution of the most extreme statistic under $H_0$. The resulting critical value $c_\alpha$ is consequently more conservative. This FWER-controlled procedure is particularly powerful when paired with our proposed statistics $t_i^{\rm (D)}$ and $t_i^{\rm (G)}$. As established in Theorems \ref{th:t} and \ref{th:margin}, our statistics for true outliers diverge and form a clear margin from the statistics of non-outliers. Therefore, even though the critical value $c_\alpha$ constructed from the FWER procedure is more conservative, we can expect that the statistics of true outliers reliably exceed this threshold, thus ensure high detection power while maintaining stringent error control. This procedure is formally described in Algorithm \ref{alg:random_rotation_fwer}.

\begin{algorithm}[t]
\caption{Outlier Detection via Random Rotation with FWER Control}
\label{alg:random_rotation_fwer}
\begin{algorithmic}[1]
    \Statex \textbf{Input}: Centered data matrix $\mathbf{X} \in \mathbb{R}^{n \times p}$, {maximum family-wise FPR} $\alpha$, number of rotations $B$.
    \Statex \textbf{Output}: Index set of declared outliers $\widehat{\mathcal{O}}$.
    \vspace{0.5em}
    
    \State Compute the statistics $\{t_1, \dots, t_n\}$ from $\mathbf{X}$.
    \State Initialize an empty set for the null distribution: $\mathcal{T}_{\text{null}} \leftarrow \emptyset$.
    
    \For{$b = 1$ to $B$}
        \State Generate a random orthogonal matrix $\mathbf{H}_b$.
        \State Compute the rotated data matrix $\mathbf{X}_b = \mathbf{H}_b\mathbf{X}$.
        \State Compute the statistics $\{t_{1,b}, \dots, t_{n,b}\}$ from $\mathbf{X}_b$.
        \State Find the maximum statistic: $t_{\max, b} = \max_{i} \{t_{i,b}\}$.
        \State Update the null distribution: $\mathcal{T}_{\text{null}} \leftarrow \mathcal{T}_{\text{null}} \cup \{t_{\max, b}\}$.
    \EndFor
    
    \State Determine the critical value $c_\alpha$ as the $(1-\alpha)$-th quantile of $\mathcal{T}_{\text{null}}$.
    \State Identify the outlier index set $\widehat{\mathcal{O}} = \{i \mid t_i > c_\alpha\}$.
    \State \textbf{Return} $\widehat{\mathcal{O}}$.
\end{algorithmic}
\end{algorithm}


\section{Simulation experiment}\label{sec:sim}

We conduct a simulation study to evaluate the empirical performance of our proposed outlier detection procedures. Our proposed procedures, denoted as DOD1, DOD2, DOD3 and DOG1, DOG2, DOG3, are based on implementing two different statistics $t_i^{\rm (D)}$ for DOD and $t_i^{\rm (G)}$ for DOG, with Algorithms \ref{alg:clustering}, \ref{alg:random_rotation_pooled}, and \ref{alg:random_rotation_fwer}, respectively. We benchmark their performance against three competing methods: Subspace Rotation-based outlier detection by \cite{chung2021subspace} (SRout), Minimum Diagonal Product by \cite{ro2015outlier} (RMDP), and Principal Component-based outlier detection by \cite{filzmoser2008outlier} (PCout). We implemented all competing methods with their default parameters as provided by the original authors.

We simulate a data matrix $\mathbf{X}$ of size $n \times p$ with $n=30$ and $p=500$ containing $n_{\text{out}}$ outliers. The generation of the $(n - n_{\text{out}})$ non-outlier observations depends on the specified structure. For the Identity (ID) and Auto-Regressive (AR) structures, the non-outliers are drawn from a multivariate normal distribution $\mathcal{N}(\mathbf{0}, \boldsymbol{\Sigma}_{\text{in}})$, where $\boldsymbol{\Sigma}_{\text{in}} = \mathbf{I}_p$ for the ID structure, and $[\boldsymbol{\Sigma}_{\text{in}}]_{j,k} = 0.7^{|j-k|}$ for the AR structure. For the Moving Average (MA) structure, non-outliers are generated directly from the process $X_j = \frac{\sum_{l=1}^{L} \eta_l Z_{j+l-1}}{(\sum_{l=1}^L \eta_l^2)^{1/2}}$ for $j=1,\dots,p$, where the $Z_k$ are independent standard normal variables, the coefficients $\eta_l$ are drawn from a uniform distribution $\mathcal{U}(0,1)$, and $L=\lfloor\sqrt{p}\rfloor$. In contrast, each outlier is independently generated from $\mathcal{N}(p^{s_{\mu}} \mathbf{u}/\|\mathbf{u}\|_2, s_{\sigma}\mathbf{I}_p)$, with elements of $\mathbf{u}$ drawn independently from $\mathcal{U}(0,1)$. The parameter $s_{\mu}$ controls the mean shift magnitude, while $s_{\sigma}$ scales the outlier covariance. Our simulations include scenarios with no outliers ($n_{\text{out}}=0$), as well as with $n_{\text{out}}=3$ under varying outlier magnitudes determined by $(s_{\mu}, s_{\sigma})$ pairs of $(0.5, 1.0)$, $(0.5, 0.5)$, and $(0.25, 0.25)$.

The tuning parameters for our proposed procedures are set as follows. For the clustering-based method detailed in Algorithm~\ref{alg:clustering}, we use k-means for clustering with $k=2$ and set the maximum allowable proportion of outliers to $\alpha = 0.3$. The gap threshold $c$ is chosen to align with the asymptotic behavior of the test statistics as stated in Corollary~\ref{cor:margin}. Specifically, we set $c = 0.1\sqrt{pn}$ for DOD1 and $c = 0.1p\sqrt{n}$ for DOG1. For the random rotation methods detailed in Algorithms \ref{alg:random_rotation_pooled} and \ref{alg:random_rotation_fwer}, we generate $B=300$ randomly rotated datasets. Further, we use $\alpha = 0.05$ for DOD2 and DOG2, and $\alpha = 0.7$ for DOD3 and DOG3.

\begin{table}
\footnotesize
\centering
\caption{Summary of simulation experiment under $n_{\rm out}=3$.}
\label{tab:sim1}
\begin{tabular}{llccccccccc}
\toprule
\multirow{2}{*}{\textbf{$(s_\mu,s_\sigma)$}} & \multirow{2}{*}{\textbf{Method}} & \multicolumn{3}{c}{\textbf{ID}} & \multicolumn{3}{c}{\textbf{AR}} & \multicolumn{3}{c}{\textbf{MA}} \\
\cmidrule(lr){3-5} \cmidrule(lr){6-8} \cmidrule(lr){9-11}
& & TPR & FPR & FWFP & TPR & FPR & FWFP & TPR & FPR & FWFP \\
\midrule
\multirow{9}{*}{{(0.5, 1.0)}}
& DOD1  & 1.000 & 0.000 & 0.000 & 1.000 & 0.000 & 0.001 & 0.996 & 0.006 & 0.141 \\
& DOD2  & 1.000 & 0.000 & 0.000 & 1.000 & 0.002 & 0.065 & 1.000 & 0.020 & 0.423 \\
& DOD3  & 1.000 & 0.000 & 0.000 & 1.000 & 0.001 & 0.024 & 1.000 & 0.011 & 0.248 \\ [3pt]
& DOG1  & 1.000 & 0.000 & 0.000 & 1.000 & 0.000 & 0.000 & 0.831 & 0.006 & 0.132 \\
& DOG2  & 1.000 & 0.000 & 0.000 & 1.000 & 0.000 & 0.000 & 0.981 & 0.002 & 0.041 \\
& DOG3  & 1.000 & 0.000 & 0.000 & 1.000 & 0.000 & 0.000 & 0.912 & 0.001 & 0.021 \\ [3pt]
& SRout & 1.000 & 0.007 & 0.176 & 1.000 & 0.007 & 0.167 & 1.000 & 0.007 & 0.164 \\
& RMDP  & 1.000 & 0.159 & 0.989 & 1.000 & 0.126 & 0.948 & 1.000 & 0.117 & 0.934 \\
& PCout & 1.000 & 0.047 & 0.692 & 0.996 & 0.055 & 0.763 & 0.793 & 0.081 & 0.860 \\
\midrule
\multirow{9}{*}{{(0.5, 0.5)}}
& DOD1  & 1.000 & 0.000 & 0.000 & 0.993 & 0.002 & 0.061 & 0.634 & 0.019 & 0.332 \\
& DOD2  & 1.000 & 0.000 & 0.001 & 1.000 & 0.004 & 0.096 & 0.881 & 0.020 & 0.420 \\
& DOD3  & 1.000 & 0.000 & 0.000 & 1.000 & 0.002 & 0.042 & 0.732 & 0.011 & 0.255 \\ [3pt]
& DOG1  & 1.000 & 0.000 & 0.000 & 1.000 & 0.000 & 0.001 & 0.833 & 0.006 & 0.133 \\
& DOG2  & 1.000 & 0.000 & 0.000 & 1.000 & 0.000 & 0.000 & 1.000 & 0.005 & 0.134 \\
& DOG3  & 1.000 & 0.000 & 0.000 & 1.000 & 0.000 & 0.000 & 0.992 & 0.002 & 0.057 \\ [3pt]
& SRout & 1.000 & 0.007 & 0.176 & 0.941 & 0.007 & 0.169 & 0.480 & 0.007 & 0.166 \\
& RMDP  & 1.000 & 0.161 & 0.985 & 1.000 & 0.127 & 0.950 & 0.965 & 0.118 & 0.926 \\
& PCout & 1.000 & 0.045 & 0.677 & 0.977 & 0.062 & 0.792 & 0.597 & 0.092 & 0.859 \\
\midrule
\multirow{9}{*}{{(0.25, 0.25)}}
& DOD1  & 1.000 & 0.000 & 0.000 & 1.000 & 0.000 & 0.004 & 0.958 & 0.019 & 0.355 \\
& DOD2  & 1.000 & 0.009 & 0.212 & 1.000 & 0.046 & 0.716 & 1.000 & 0.079 & 0.905 \\
& DOD3  & 1.000 & 0.005 & 0.126 & 1.000 & 0.037 & 0.613 & 1.000 & 0.063 & 0.827 \\ [3pt]
& DOG1  & 0.000 & 0.000 & 0.000 & 0.000 & 0.000 & 0.000 & 0.000 & 0.002 & 0.033 \\
& DOG2  & 0.000 & 0.000 & 0.000 & 0.000 & 0.022 & 0.464 & 0.033 & 0.064 & 0.922 \\
& DOG3  & 0.000 & 0.000 & 0.000 & 0.000 & 0.014 & 0.323 & 0.014 & 0.045 & 0.792 \\ [3pt]
& SRout & 0.000 & 0.119 & 0.952 & 0.000 & 0.088 & 0.915 & 0.000 & 0.063 & 0.808 \\
& RMDP  & 0.000 & 0.048 & 0.671 & 0.000 & 0.100 & 0.911 & 0.000 & 0.165 & 0.981 \\
& PCout & 0.990 & 0.055 & 0.745 & 0.957 & 0.050 & 0.702 & 0.746 & 0.071 & 0.814 \\
\bottomrule
\end{tabular}
\end{table}

Table~\ref{tab:sim1} summarizes the simulation results from 1000 replicates for scenarios with three outliers ($n_{\text{out}}=3$). We assess the performance of each method using three metrics: the average True Positive Rate (TPR), representing the proportion of true outliers correctly identified; the average False Positive Rate (FPR), representing the proportion of non-outliers incorrectly flagged as outliers; and the Family-Wise False Positive rate (FWFP), which is the proportion of simulation replicates containing at least one incorrect non-outlier flagging.

The results in Table~\ref{tab:sim1} demonstrate that our proposed DOD and DOG methods outperform the competitors by providing a superior balance between detection power and error control. Across nearly all settings, our method DOD achieves a perfect or near-perfect TPR of 1.000, successfully identifying all true outliers. Crucially, our methods accomplish this detection power while maintaining an FPR at or very near zero, indicating robust control over false discoveries. A critical distinction is observed between the two rotation-based methods DOD2 and DOD3, particularly in their control of FWFP. DOD3 aims to control the FWER by its design, and the results confirm its success; DOD3 consistently yields a substantially lower FWFP than DOD2, without compromising its TPR. In contrast, while the competing methods (SRout, RMDP, and PCout) can also exhibit high TPR, they do so at the cost of inflated error rates. Their FPR is consistently higher, and their FWFP frequently exceeds 0.5 and often approaches 1.0, implying that they incorrectly flag non-outliers in the majority of replicates. Moreover, even when outliers were subtle with outlyingness magnitude $(s_{\mu}, s_{\sigma})=(0.25, 0.25)$, DOD1--3 maintained TPR at perfect or near 1.000, while SRout and RMDP failed completely, and PCout's detection power was diminished.

\begin{table}
\footnotesize
\centering
\caption{Summary of simulation experiment under $n_{\rm out}=0$.}\label{tab:sim2}
\begin{tabular}{lcccccc}

\toprule
\multirow{2}*{\textbf{Method}}
& \multicolumn{2}{c}{\textbf{ID}}
& \multicolumn{2}{c}{\textbf{AR}}
& \multicolumn{2}{c}{\textbf{MA}} \\
\cmidrule(lr){2-3} \cmidrule(lr){4-5} \cmidrule(lr){6-7}
& FPR & FWFP & FPR & FWFP & FPR & FWFP \\
\midrule
DOD1  & 0.001 & 0.027 & 0.005 & 0.108 & 0.016 & 0.246 \\
DOD2  & 0.009 & 0.224 & 0.033 & 0.634 & 0.045 & 0.755 \\
DOD3  & 0.004 & 0.105 & 0.026 & 0.538 & 0.035 & 0.655 \\ [3pt]
DOG1  & 0.000 & 0.000 & 0.000 & 0.000 & 0.001 & 0.029 \\
DOG2  & 0.000 & 0.000 & 0.020 & 0.491 & 0.050 & 0.877 \\
DOG3  & 0.000 & 0.000 & 0.012 & 0.324 & 0.034 & 0.709 \\ [3pt]
SRout & 0.010 & 0.237 & 0.009 & 0.244 & 0.010 & 0.229 \\
RMDP  & 0.158 & 0.985 & 0.141 & 0.975 & 0.129 & 0.935 \\
PCout & 0.121 & 0.961 & 0.116 & 0.961 & 0.123 & 0.962 \\
\bottomrule

\end{tabular}
\end{table}

Table~\ref{tab:sim2} presents the simulation results where no outliers are present in the data ($n_{\text{out}}=0$). In this setting, the ideal method should refrain from declaring outliers, thus achieving FPR or FWFP below the prespecified maximum FPR $\alpha$. The results demonstrate the superiority of our proposed methods in controlling error. Our methods exhibit outstanding performance, maintaining a nearly perfect FPR of 0.000 under both ID and AR structures, thus making almost no incorrect outlier flagging. In addition, consistently with tuning parameters, DOD2 and DOG2 control FPR below its prespcified level $\alpha=0.05$, and DOD3 and DOG3 control FWFP around or below its prespcified level $\alpha=0.7$. These results empirically confirm that our FWER-controlling algorithms (DOD3, DOG3) provide reliable control over family-wise false positives, making them suitable for robust error control. In contrast, RMDP and PCout perform poorly with FWFP consistently above 0.9, indicating that they incorrectly declare outliers in almost every single replicate.


\section{Real data application}\label{sec:real}
\subsection{Microarray gene expression}\label{sec:real-gene}

To further evaluate the empirical performance of our proposed methods against competing methods, we analyze the lymphoma microarray gene expression dataset \citep{dettling2004bagboosting}. Described in \cite{alizadeh2000distinct}, the dataset contains expression measurements of $p=4026$ genes for $n=62$ samples. The samples belong to three lymphoma types, where the largest class, Diffuse Large B-Cell Lymphoma, consists of 42 samples, which will be designated as inliers. The remaining 20 samples from the other two classes will serve as a pool of potential outliers.

Specifically, we designed two experimental scenarios to assess the methods' performance:
\begin{enumerate}
    \item Contaminated case with $n_{\text{out}} = 2$: In each replication, the dataset was constructed using all 42 inlier samples and 2 outlier samples randomly drawn from the pool of 20. This scenario tests the methods' ability to correctly identify true outliers while avoiding false positives. We performed 200 replications.
    \item Null case with $n_{\text{out}} = 0$: This dataset consisted solely of the 42 inlier samples. This scenario is designed to evaluate the methods' control over the false positive rate when no true outliers are present.
\end{enumerate}

The performance of each method was measured using the average True Positive Rate (TPR), False Positive Rate (FPR), and Family-Wise False Positive Rate (FWFP), which were defined the same as in Section \ref{sec:sim}. Our proposed procedures, DOD1--3 based on $t_i^{\rm (D)}$ and DOG1--3 based on $t_i^{\rm (G)}$, were implemented with parameters $B=300$ for the random rotation algorithms, and $\alpha$ same as in Section \ref{sec:sim}. The competing methods SRout, RMDP, and PCout were implemented again using their default parameters as specified by the original authors.

\begin{table}
    \footnotesize
    \centering
    \caption{Summary of microarray gene expression analysis under {$n_{\text{out}} = 2$}.}
    \label{tab:lymphoma2}
    \begin{tabular}{lrrr}
    \toprule
    Method &   TPR &   FPR &  FWFP \\
    \midrule
    DOD1   & 1.000 & 0.020 & 0.320 \\
    DOD2   & 1.000 & 0.026 & 0.990 \\
    DOD3   & 1.000 & 0.000 & 0.000 \\ [3pt]
    DOG1   & 0.907 & 0.061 & 0.465 \\
    DOG2   & 0.823 & 0.000 & 0.005 \\
    DOG3   & 0.700 & 0.000 & 0.000 \\ [3pt]
    SRout  & 1.000 & 0.081 & 1.000 \\
    RMDP   & 0.880 & 0.017 & 0.725 \\
    PCout  & 0.178 & 0.137 & 1.000 \\
    \bottomrule
    \end{tabular}
\end{table}

\begin{table}
    \footnotesize
    \centering
    \caption{Summary of microarray gene expression analysis under $n_{\rm out}=0$.}
    \label{tab:lymphoma0}
    \begin{tabular}{lrr}
    \toprule
    Method &   FPR &  FWFP \\
    \midrule
    DOD1   & 0.095 & 1.000 \\
    DOD2   & 0.024 & 1.000 \\
    DOD3   & 0.000 & 0.015 \\ [3pt]
    DOG1   & 0.000 & 0.000 \\
    DOG2   & 0.024 & 1.000 \\
    DOG3   & 0.000 & 0.000 \\ [3pt]
    SRout  & 0.133 & 1.000 \\
    RMDP   & 0.017 & 0.720 \\
    PCout  & 0.143 & 1.000 \\
    \bottomrule
    \end{tabular}
\end{table}

The summarized results for both scenarios are presented in Tables \ref{tab:lymphoma2} and \ref{tab:lymphoma0}. In the contaminated case with {$n_{\text{out}} = 2$}, as shown in Table \ref{tab:lymphoma2}, our proposed methods demonstrated excellent detection power. The key distinction among the methods lies in their control of false positives. {DOD1--3} performed exceptionally well, combining perfect TPR with very low FPR, with DOD3 in particular maintaining a perfect record of zero false positives (FPR = 0 and FWFP = 0). In addition, DOG1-3 achieved better balances between TPR and FPR than the competing methods. In contrast, while a competing method SRout showed a perfect TPR, it came at the cost of a FWFP of 1.0, suggesting it always flags inliers. PCout performed poorly in terms of both detection power {(TPR = 0.178)} and false positive control (FWFP = 1.0).

In the null case with $n_{\text{out}} = 0$, as shown in Table \ref{tab:lymphoma0}, the superiority of the {our proposed} methods was evident. DOG1 and DOG3 exhibited perfect control over false positives, achieving a perfect FWFP of 0. {DOD3} performed robustly with a near-zero FWFP of {0.015.} The remaining methods, including our DOD1, DOD2, DOG2 and all three competitors, struggled significantly in this scenario, with FWFP values ranging from 0.72 to a worst 1.0. This shows that these methods are prone to flagging outliers even when none exist.


\subsection{Human face image}\label{sec:real-face}

As a second real data application, we analyze the Olivetti Research Laboratory face image dataset \citep{samaria1994parameterisation}. This dataset comprises 400 grayscale images of 40 distinct individuals, with 10 different images per person capturing various facial expressions and lighting conditions. Each image consists of $112 \times 92$ pixels, resulting in a high-dimensional feature with $p=10304$ variables.

The experimental design was structured as follows. For each of the 40 individuals, their 10 images were designated as the inlier group. The remaining 390 images from the other 39 individuals served as a pool of potential outliers. Specifically, we investigated two scenarios:
\begin{enumerate}
    \item Contaminated case with $n_{\rm out} = 1$: The dataset was constructed with 10 inlier images from one individual, and one outlier image randomly selected from the pool of 390.
    \item Null case with $n_{\rm out} = 0$: The dataset consisted solely of the 10 inlier images from one individual.
\end{enumerate}
This entire process was repeated for each of the 40 individuals, and for each individual, the experiment was replicated 5 times, leading to a total of 200 independent runs for each scenario. Performance was evaluated using the average TPR, FPR, and FWFP, as defined in Section \ref{sec:sim}. For this experiment, $\alpha$ for Algorithm \ref{alg:random_rotation_pooled} was set to 0.1 to reflect the small sample size $n=10$ of the inlier group.

\begin{table}
    \footnotesize
    \centering
    \caption{Summary of human face image analysis under $n_{\rm out}=1$.}
    \label{tab:face1}
    \begin{tabular}{lrrr}
    \toprule
    Method &   TPR &   FPR &  FWFP \\
    \midrule
    DOD1   & 0.970 & 0.019 & 0.125 \\
    DOD2   & 0.975 & 0.088 & 0.330 \\
    DOD3   & 0.975 & 0.086 & 0.315 \\ [3pt]
    DOG1   & 0.735 & 0.038 & 0.205 \\
    DOG2   & 0.130 & 0.062 & 0.195 \\
    DOG3   & 0.180 & 0.062 & 0.200 \\ [3pt]
    SRout  & 0.980 & 0.118 & 0.785 \\
    RMDP   & 0.905 & 0.060 & 0.370 \\
    PCout  & 0.845 & 0.231 & 0.885 \\
    \bottomrule
    \end{tabular}
\end{table}

\begin{table}
    \footnotesize
    \centering
    \caption{Summary of human face image analysis under $n_{\rm out}=0$.}
    \label{tab:face0}
    \begin{tabular}{lrr}
    \toprule
    Method &   FPR &  FWFP \\
    \midrule
    DOD1   & 0.132 & 0.750 \\
    DOD2   & 0.260 & 0.950 \\
    DOD3   & 0.264 & 0.945 \\ [3pt]
    DOG1   & 0.132 & 0.600 \\
    DOG2   & 0.158 & 0.635 \\
    DOG3   & 0.170 & 0.675 \\ [3pt]
    SRout  & 0.178 & 0.905 \\
    RMDP   & 0.065 & 0.350 \\
    PCout  & 0.283 & 0.900 \\
    \bottomrule
    \end{tabular}
\end{table}

The summarized results are presented in Tables \ref{tab:face1} and \ref{tab:face0}. In the contaminated case with $n_{\rm out} = 1$, the results presented in Table \ref{tab:face1} show that our proposed methods DOD1--3 and the competing method SRout exhibited the highest detection power with TPRs around 0.975. Among the top-performing methods, DOD1--3 provided better FPR and FWFP control than SRout. RMDP ranked just below this group, showing high detection power and reasonable FPR and FWFP control.

In the null case with $n_{\rm out} = 0$, where the focus is on controlling false discoveries, Table \ref{tab:face0} shows that RMDP achieved the best performance with the lowest FPR and FWFP values. Our proposed methods DOD1 and DOG1--3 also performed reasonably well. The other procedures struggled to control false positives, yielding FWFP values $0.9$ or higher.


\section{Conclusion}\label{sec:con}

In this paper, we proposed two statistics for outlier detection in high-dimensional data. These statistics leverage pairwise distances and inner products to capture an observation's relational dissimilarities. We provided a theoretical foundation for these statistics, demonstrating that as the dimension increases, a non-vanishing margin asymptotically separates outliers from non-outliers. Based on this theoretical guarantee, we developed three practical detection procedures: a clustering-based method and two non-parametric tests based on random rotation, one of which offers robust control over the family-wise error rate. Our simulation studies and real data applications demonstrated that the proposed methods achieve a balance of high detection power and stringent control over false discoveries.

Several avenues for future research remain. First, while our work establishes the asymptotic properties of the statistics, an investigation into their finite dimension behavior under less restrictive assumptions would be a valuable theoretical extension. Second, the current framework is presented based on Euclidean distances and dot products; it could be extended to incorporate other dissimilarity metrics to handle a wider range of data types and outlier mechanisms. Finally, while the random rotation tests are powerful, they may be computationally intensive. Developing faster, deterministic approximations or exploring more computationally efficient resampling schemes could enhance the practical applicability of our methods for massive datasets.


\section*{Supplementary material}

The supplementary material includes all of the technical details.

\section*{Data availability}

The lymphoma microarray gene expression dataset analyzed in Section \ref{sec:real-gene} is available at the R package \texttt{spls}, and the Olivetti Research Laboratory face image dataset analyzed in Section \ref{sec:real-face} is available at the following URL: https://www.kaggle.com/code/serkanpeldek/face-recognition-on-olivetti-dataset.

\section*{Acknowledgments}

This work was supported by the 2024 Research Fund of the University of Seoul for Seong-ho Lee, and the National Research Foundation of Korea (NRF) grant (No. RS-2025-16069168) and the IITP-ICAN grant (No. RS-2023-00259934) funded by the Korea government (MSIT) for Yongho Jeon.




\bibliographystyle{agsm}
\bibliography{HDLSSoutlier}

@article{ahn2010maximal,
  title={The maximal data piling direction for discrimination},
  author={Ahn, Jeongyoun and Marron, JS},
  journal={Biometrika},
  volume={97},
  number={1},
  pages={254--259},
  year={2010},
  publisher={Oxford University Press}
}

@article{ahn2012clustering,
  title={Clustering high dimension, low sample size data using the maximal data piling distance},
  author={Ahn, Jeongyoun and Lee, Myung Hee and Yoon, Young Joo},
  journal={Statistica Sinica},
  pages={443--464},
  year={2012},
  publisher={JSTOR}
}

@article{chung2021subspace,
  title={Subspace rotations for high-dimensional outlier detection},
  author={Chung, Hee Cheol and Ahn, Jeongyoun},
  journal={Journal of Multivariate Analysis},
  volume={183},
  pages={104713},
  year={2021},
  publisher={Elsevier}
}

@article{alizadeh2000distinct,
  title={Distinct types of diffuse large B-cell lymphoma identified by gene expression profiling},
  author={Alizadeh, Ash A and Elsen, Michael B and Davis, R Eric and Ma, Chi and others},
  journal={Nature},
  volume={403},
  number={6769},
  pages={503},
  year={2000},
  publisher={Nature Publishing Group}
}

@article{dettling2004bagboosting,
  title={BagBoosting for tumor classification with gene expression data},
  author={Dettling, Marcel},
  journal={Bioinformatics},
  volume={20},
  number={18},
  pages={3583--3593},
  year={2004},
  publisher={Oxford University Press}
}

@article{filzmoser2008outlier,
  title={Outlier identification in high dimensions},
  author={Filzmoser, Peter and Maronna, Ricardo and Werner, Mark},
  journal={Computational Statistics \& Data Analysis},
  volume={52},
  number={3},
  pages={1694--1711},
  year={2008},
  publisher={Elsevier}
}

@article{hall2005geometric,
  title={Geometric representation of high dimension, low sample size data},
  author={Hall, Peter and Marron, JS and Neeman, Amnon},
  journal={Journal of the Royal Statistical Society: Series B (Statistical Methodology)},
  volume={67},
  number={3},
  pages={427--444},
  year={2005},
  publisher={Wiley Online Library}
}

@article{ro2015outlier,
  title={Outlier detection for high-dimensional data},
  author={Ro, Kwangil and Zou, Changliang and Wang, Zhaojun and Yin, Guosheng},
  journal={Biometrika},
  volume={102},
  number={3},
  pages={589--599},
  year={2015},
  publisher={Oxford University Press}
}

@article{terada2013clustering,
  title={Clustering for high-dimension, low-sample size data using distance vectors},
  author={Terada, Yoshikazu},
  journal={arXiv preprint arXiv:1312.3386},
  year={2013}
}

@inproceedings{kriegel2008angle,
  title={Angle-based outlier detection in high-dimensional data},
  author={Kriegel, Hans-Peter and Zimek, Arthur and others},
  booktitle={Proceedings of the 14th ACM SIGKDD international conference on Knowledge discovery and data mining},
  pages={444--452},
  year={2008},
  organization={ACM}
}

@inproceedings{papadimitriou2003loci,
  title={Loci: Fast outlier detection using the local correlation integral},
  author={Papadimitriou, Spiros and Kitagawa, Hiroyuki and Gibbons, Phillip B and Faloutsos, Christos},
  booktitle={Data Engineering, 2003. Proceedings. 19th International Conference on},
  pages={315--326},
  year={2003},
  organization={IEEE}
}

@inproceedings{breunig2000lof,
  title={LOF: identifying density-based local outliers},
  author={Breunig, Markus M and Kriegel, Hans-Peter and Ng, Raymond T and Sander, J{\"o}rg},
  booktitle={ACM sigmod record},
  volume={29},
  number={2},
  pages={93--104},
  year={2000},
  organization={ACM}
}

@article{chandola2009anomaly,
  title={Anomaly detection: A survey},
  author={Chandola, Varun and Banerjee, Arindam and Kumar, Vipin},
  journal={ACM computing surveys (CSUR)},
  volume={41},
  number={3},
  pages={15},
  year={2009},
  publisher={ACM}
}

@article{zimek2012survey,
  title={A survey on unsupervised outlier detection in high-dimensional numerical data},
  author={Zimek, Arthur and Schubert, Erich and Kriegel, Hans-Peter},
  journal={Statistical Analysis and Data Mining: The ASA Data Science Journal},
  volume={5},
  number={5},
  pages={363--387},
  year={2012},
  publisher={Wiley Online Library}
}

@article{mcgill1978variations,
  title={Variations of box plots},
  author={McGill, Robert and Tukey, John W and Larsen, Wayne A},
  journal={The American Statistician},
  volume={32},
  number={1},
  pages={12--16},
  year={1978},
  publisher={Taylor \& Francis Group}
}

@inproceedings{ester1996density,
  title={A density-based algorithm for discovering clusters in large spatial databases with noise.},
  author={Ester, Martin and Kriegel, Hans-Peter and Sander, J{\"o}rg and Xu, Xiaowei and others},
  booktitle={Kdd},
  volume={96},
  number={34},
  pages={226--231},
  year={1996}
}

@article{ye2001anomaly,
  title={An anomaly detection technique based on a chi-square statistic for detecting intrusions into information systems},
  author={Ye, Nong and Chen, Qiang},
  journal={Quality and Reliability Engineering International},
  volume={17},
  number={2},
  pages={105--112},
  year={2001},
  publisher={Wiley Online Library}
}

@article{blaser2016random,
  title={Random rotation ensembles},
  author={Blaser, Rico and Fryzlewicz, Piotr},
  journal={The Journal of Machine Learning Research},
  volume={17},
  number={1},
  pages={126--151},
  year={2016},
  publisher={JMLR. org}
}

@inproceedings{samaria1994parameterisation,
  title={Parameterisation of a stochastic model for human face identification},
  author={Samaria, Ferdinando S and Harter, Andy C},
  booktitle={Proceedings of 1994 IEEE workshop on applications of computer vision},
  pages={138--142},
  year={1994},
  organization={IEEE}
}


\clearpage
\pagenumbering{arabic}
\section*{Supplementary material}
\setcounter{equation}{0}\renewcommand{\theequation}{S.\arabic{equation}}
\setcounter{subsection}{0}\renewcommand{\thesubsection}{S.\arabic{subsection}}


\subsection{Proof of Theorem \ref{th:t}}\label{sec:supp-th:t}

We provide the proof for $t_i^{\rm (D)}$; the argument for $t_i^{\rm (G)}$ is analogous. The proof proceeds in two main steps. First, we establish the asymptotic limits of the component terms $\delta_{ij}^{\rm (D)}$ and the column-wise medians $\widetilde{\delta}_{.j}^{\rm (D)}$. Second, we use these limits to prove parts (i) and (ii) of the theorem.

We begin by analyzing the probabilistic limit of $\frac{1}{\sqrt{p}}\delta_{ij}^{\rm (D)}$. By definition,
\[
\frac{1}{p}\left\{\delta_{ij}^{\rm (D)}\right\}^2 = \sum_{k \neq i,j} \left(\frac{[\D]_{i,k}-[\D]_{j,k}}{\sqrt{p}}\right)^2.
\]
We consider the limit of this sum based on the nature of $\mathbf{x}_i$ and $\mathbf{x}_j$, applying the results from Lemma \ref{lem:terada}.

\begin{enumerate}
    \item \textbf{If $\mathbf{x}_i, \mathbf{x}_j \sim \mathbf{x}^{\rm (I)}$ (both non-outliers):} For any third point $\mathbf{x}_k$, the term $([\D]_{i,k}-[\D]_{j,k})/\sqrt{p}$ converges to zero in probability by Lemma \ref{lem:terada}, regardless of whether $\mathbf{x}_k$ is a non-outlier or an outlier. Thus, every term in the sum converges to zero, which implies $\frac{1}{\sqrt{p}}\delta_{ij}^{\rm (D)} \xrightarrow{p} 0$.

    \item \textbf{If $\mathbf{x}_i, \mathbf{x}_j \sim \mathbf{x}^{\rm (O)}$ (both outliers):} For any third point $\mathbf{x}_k$, the term $([\D]_{i,k}-[\D]_{j,k})/\sqrt{p}$ converges to zero by Lemma \ref{lem:terada}, as the distances from two outliers to any third point are asymptotically equivalent. This leads to $\frac{1}{\sqrt{p}}\delta_{ij}^{\rm (D)} \xrightarrow{p} 0$.

    \item \textbf{If $\mathbf{x}_i \sim \mathbf{x}^{\rm (I)}, \mathbf{x}_j \sim \mathbf{x}^{\rm (O)}$ (one non-outlier, one outlier):} The set of $\mathbf{x}_k$ consists of $(n-n_{\rm out}-1)$ non-outliers excluding $\mathbf{x}_i$, and $(n_{\rm out}-1)$ outliers excluding $\mathbf{x}_j$. The sum of squared limits is then $(n-n_{\rm out}-1)\alpha_{\rm D}^2 + (n_{\rm out}-1)\beta_{\rm D}^2=\gamma_{\rm D}^2$ by Lemma \ref{lem:terada}. It follows that $\frac{1}{\sqrt{p}}\delta_{ij}^{\rm (D)} \xrightarrow{p} \gamma_{\rm D}$.
\end{enumerate}

Next, we determine the limit of the scaled column-wise median, $\frac{1}{\sqrt{p}}\widetilde{\delta}_{.j}^{\rm (D)}$. Since we can assume $n_{\rm out} < n/2$ by the definition of outlier, the median is determined by the behavior of the non-outlier rows.

\begin{enumerate}
    \item \textbf{If $\mathbf{x}_j \sim \mathbf{x}^{\rm (I)}$:} The $j$-th column of $\bDelta_{\rm D}$ consists of a majority of $\delta_{ij}$ values where $\mathbf{x}_i \sim \mathbf{x}^{\rm (I)}$ whose scaled limit is 0, and a minority where $\mathbf{x}_i \sim \mathbf{x}^{\rm (O)}$ whose scaled limit is $\gamma_{\rm D}$. Therefore, $\frac{1}{\sqrt{p}}\widetilde{\delta}_{.j}^{\rm (D)} \xrightarrow{p} 0$.
    \item \textbf{If $\mathbf{x}_j \sim \mathbf{x}^{\rm (O)}$:} The $j$-th column consists of a majority of values whose scaled limit is $\gamma_{\rm D}$ and a minority whose scaled limit is 0. Therefore, $\frac{1}{\sqrt{p}}\widetilde{\delta}_{.j}^{\rm (D)} \xrightarrow{p} \gamma_{\rm D}$.
\end{enumerate}

With these component limits, we now prove the main statements.
\begin{enumerate}[label=(\roman*)]
    \item \textbf{If $\mathbf{x}_i \sim \mathbf{x}^{\rm (I)}$:} We examine the limit of $\frac{1}{p}\{t_i^{\rm (D)}\}^2 = \sum_{j=1}^n \left\{\frac{\delta_{ij}^{\rm (D)}}{\sqrt{p}} - \frac{\widetilde{\delta}_{.j}^{\rm (D)}}{\sqrt{p}}\right\}^2$.
    \begin{itemize}
        \item For terms where $\mathbf{x}_j \sim \mathbf{x}^{\rm (I)}$, the squared difference converges to $(0-0)^2 = 0$.
        \item For terms where $\mathbf{x}_j \sim \mathbf{x}^{\rm (O)}$, the squared difference converges to $(\gamma_{\rm D}-\gamma_{\rm D})^2 = 0$.
    \end{itemize}
    Since every term in the sum converges to zero, $\frac{1}{p}\{t_i^{\rm (D)}\}^2 \xrightarrow{p} 0$, which implies $\frac{1}{\sqrt{p}}t_i^{\rm (D)} \xrightarrow{p} 0$.

    \item \textbf{If $\mathbf{x}_i \sim \mathbf{x}^{\rm (O)}$:} We analyze the same sum $\frac{1}{p}\{t_i^{\rm (D)}\}^2 = \sum_{j=1}^n \left\{\frac{\delta_{ij}^{\rm (D)}}{\sqrt{p}} - \frac{\widetilde{\delta}_{.j}^{\rm (D)}}{\sqrt{p}}\right\}^2$.
    \begin{itemize}
        \item For the $(n-n_{\rm out})$ terms where $\mathbf{x}_j \sim \mathbf{x}^{\rm (I)}$, the squared difference converges to $(\gamma_{\rm D}-0)^2 = \gamma_{\rm D}^2$.
        \item For the $n_{\rm out}$ terms where $\mathbf{x}_j \sim \mathbf{x}^{\rm (O)}$, the squared difference converges to $(0-\gamma_{\rm D})^2 = \gamma_{\rm D}^2$.
    \end{itemize}
    Every one of the $n$ terms in the sum converges to $\gamma_{\rm D}^2$. Therefore, the sum of the limits is $n\gamma_{\rm D}^2$:
    \[
    \frac{1}{p}\{t_i^{\rm (D)}\}^2 = \sum_{j=1}^n \left\{\frac{\delta_{ij}^{\rm (D)}}{\sqrt{p}} - \frac{\widetilde{\delta}_{.j}^{\rm (D)}}{\sqrt{p}}\right\}^2 \xrightarrow{p} n\gamma_{\rm D}^2.
    \]
    Diving both sides by $n$, we obtain the final result:
    \[
    \frac{1}{np}(t_i^{\rm (D)})^2 \xrightarrow{p} \gamma_{\rm D}^2 \quad \implies \quad \frac{1}{\sqrt{pn}}t_i^{\rm (D)} \xrightarrow{p} \gamma_{\rm D}.
    \]
\end{enumerate}
This completes the proof.
\hfill$\blacksquare$


\subsection{Proof of Theorem \ref{th:margin}}\label{sec:supp-th:margin}

We provide the proof for the distance-based statistic $t_i^{\rm (D)}$; the proof for $t_i^{\rm (G)}$ follows analogously. The proof consists of establishing the limits for the maximum of the scaled non-outlier statistics and the minimum of the scaled outlier statistics separately, and then combining them. Let $M_p^{(\mathcal{I})} = \max_{i \in \mathcal{I}} \frac{t_i^{\rm (D)}}{\sqrt{pn}}$ and $m_p^{(\mathcal{O})} = \min_{i \in \mathcal{O}} \frac{t_i^{\rm (D)}}{\sqrt{pn}}$.

First, we show that $M_p^{(\mathcal{I})} \xrightarrow{p} 0$.
From Theorem \ref{th:t}(i), we know that for any individual non-outlier $i \in \mathcal{I}$, $\frac{t_i^{\rm (D)}}{\sqrt{p}} \xrightarrow{p} 0$. Rescaling this term gives:
\[
\frac{t_i^{\rm (D)}}{\sqrt{pn}} = \frac{1}{\sqrt{n}}\left(\frac{t_i^{\rm (D)}}{\sqrt{p}}\right) \xrightarrow{p} 0.
\]
To show that the maximum also converges to zero, we use the union bound for any $\epsilon > 0$:
\[
\Pr\{M_p^{(\mathcal{I})} \geq \epsilon\} = \Pr\left[\bigcup_{i \in \mathcal{I}} \left\{\frac{t_i^{\rm (D)}}{\sqrt{pn}} \geq \epsilon\right\}\right] \leq \sum_{i \in \mathcal{I}} \Pr\left\{\frac{t_i^{\rm (D)}}{\sqrt{pn}} \geq \epsilon\right\}.
\]
Since the number of non-outliers, $|\mathcal{I}|$, is a fixed finite number and each term in the sum converges to 0 as $p \to \infty$, their sum also converges to 0. Thus, $M_p^{(\mathcal{I})} \xrightarrow{p} 0$.

Next, we show that $m_p^{(\mathcal{O})} \xrightarrow{p} \gamma_{\rm D}$.
From Theorem \ref{th:t}(ii), for any individual outlier $i \in \mathcal{O}$, we have $\frac{t_i^{\rm (D)}}{\sqrt{pn}} \xrightarrow{p} \gamma_{\rm D}$. To show that the minimum converges to the same limit, we consider for any $\epsilon > 0$:
\[
\Pr\{|m_p^{(\mathcal{O})} - \gamma_{\rm D}| \geq \epsilon\} \leq \Pr\{m_p^{(\mathcal{O})} \geq \gamma_{\rm D} + \epsilon\} + \Pr\{m_p^{(\mathcal{O})} \leq \gamma_{\rm D} - \epsilon\}.
\]
The first term $\Pr\{m_p^{(\mathcal{O})} \geq \gamma_{\rm D} + \epsilon\}$ is less than or equal to $\Pr\left\{\frac{t_j^{\rm (D)}}{\sqrt{pn}} \geq \gamma_{\rm D} + \epsilon\right\}$ for any single $j \in \mathcal{O}$, which converges to 0. For the second term, we again use the union bound:
\[
\Pr\{m_p^{(\mathcal{O})} \leq \gamma_{\rm D} - \epsilon\} = \Pr\left[\bigcup_{i \in \mathcal{O}} \left\{\frac{t_i^{\rm (D)}}{\sqrt{pn}} \leq \gamma_{\rm D} - \epsilon\right\}\right] \leq \sum_{i \in \mathcal{O}} \Pr\left\{\frac{t_i^{\rm (D)}}{\sqrt{pn}} \leq \gamma_{\rm D} - \epsilon\right\}.
\]
Since $|\mathcal{O}|$ is a fixed finite number and each term in the sum converges to 0, the sum converges to 0. Therefore, $\Pr\{|m_p^{(\mathcal{O})} - \gamma_{\rm D}| \geq \epsilon\} \to 0$, which proves $m_p^{(\mathcal{O})} \xrightarrow{p} \gamma_{\rm D}$.

Combining the results, we get:
\[
m_p^{(\mathcal{O})} - M_p^{(\mathcal{I})} = \min_{i \in \mathcal{O}} \frac{t_i^{\rm (D)}}{\sqrt{pn}} - \max_{i \in \mathcal{I}} \frac{t_i^{\rm (D)}}{\sqrt{pn}} \xrightarrow{p} \gamma_{\rm D} - 0 = \gamma_{\rm D}.
\]
This completes the proof.
\hfill$\blacksquare$


\end{document}